\documentclass[letterpaper,journal,draft,onecolumn]{IEEEtran}%
\usepackage{amsfonts}
\usepackage{amssymb}
\usepackage{amsmath}
\usepackage{graphicx}%
\providecommand{\U}[1]{\protect\rule{.1in}{.1in}}

\ifx\pdfoutput\relax\let\pdfoutput=\undefined\fi
\newcount\msipdfoutput
\ifx\pdfoutput\undefined\else
\ifcase\pdfoutput\else
\ifx\paperwidth\undefined\else
\ifdim\paperheight=0pt\relax\else\pdfpageheight\paperheight\fi
\ifdim\paperwidth=0pt\relax\else\pdfpagewidth\paperwidth\fi
\fi\fi\fi
\begin{document}

\title{An Analysis of Phase Synchronization Mismatch Sensitivity for Coherent MIMO
Radar Systems}
\author{Hana Godrich$^{\circ}$,~Alexander M. Haimovich$^{\circ}$,~and H. Vincent
Poor$^{\dagger}$\\$^{\circ}$New Jersey Institute of Technology, Newark, NJ 07102\\$^{\dagger}$Princeton University, Princeton, NJ 08544\thanks{The research was
supported in part by the office of Naval Research under grant N00014-09-0342.
}}
\maketitle

\begin{abstract}
In this study, the hybrid Cramer-Rao bound (CRB) is developed for target
localization, to establish the sensitivity of the estimation mean-square error
(MSE) to the level of phase synchronization mismatch in coherent
Multiple-Input Multiple-Output (MIMO) radar systems with widely distributed
antennas. The lower bound on the MSE is derived for the joint estimation of
the vector of unknown parameters, consisting of the target location and the
mismatch of the allegedly known system parameters, i.e., phase offsets at the
radars. Synchronization errors are modeled as being random and Gaussian. A
closed-form expression for the hybrid CRB is derived for the case of
orthogonal waveforms.The bound on the target localization MSE is expressed as
the sum of two terms; the first represents the CRB with no phase mismatch, and
the second captures the mismatch effect. The latter is shown to depend on the
phase error variance, the number of mismatched transmitting and receiving
sensors and the system geometry. For a given phase synchronization error
variance, this expression offers the means to analyze the achievable
localization accuracy. Alternatively, for a predetermined localization MSE
target value, the derived expression may be used to determine the necessary
phase synchronization level in the distributed system.

\end{abstract}


\section{Introduction}

\label{Section:Introduction}

Improvement in target parameter estimation capabilities is a primary advantage
of MIMO radar systems \cite{Fishler04}-\cite{Godrich09}. In particular, target
localization with coherent MIMO radar systems, utilizing widely distributed
antennas, offers significant advantages \cite{Godrich09}. Typically,
performance analysis of system parameter estimation problems is based on the
derivation of the Cramer-Rao bound (CRB), which sets a lower bound on the
estimation MSE for unbiased estimators \cite{Poor}. Such an evaluation is
provided in \cite{HaimovichBook}, \cite{Godrich09} for coherent MIMO radar
systems, demonstrating a localization accuracy advantage, inversely
proportional to the signal carrier frequency. In addition, a spatial advantage
of the order of the product of the number of transmit and receive radars is
also incorporated in the CRB.

This performance gain comes with the challenge of attaining phase
synchronization in a distributed system. Errors introduced to the system
parameters by phase synchronization mismatch, will result in parameter
estimation mean-square error (MSE) degradation and bias. In this work, the
\emph{hybrid} CRB (HCRB) is used to test the sensitivity of the target
localization MSE to phase errors. The HCRB takes into account deterministic
unknown parameters, such as the target location, as well as random parameters,
phase calibration errors, in this case. This method has been applied to
passive source localization \cite{Rockah87}, \cite{Messer88} for the problem
of source bearing and range estimation with uncertainty in the sensors'
locations or phase synchronization errors.

In this work, the HCRB is derived for coherent MIMO radars, with phase
synchronization errors. A closed-form expression for the HCRB for the target's
location $\left(  x,y\right)  $ is derived, providing the means to assess the
effects of phase errors on the localization accuracy. The effect of the number
of radars, their geometric layout, and the phase mismatch MSE is incorporated
in the HCRB terms.

The paper is organized as follows: brief theoretical background is provided in
Section \ref{Section:Background}. The system model is introduced in Section
\ref{Section:HCRB}, and the HCRB\ on the targets localization estimation
errors is derived. Numerical examples are presented in Section
\ref{Section:NumericalAnalysis}. Finally, Section \ref{Section:Conclusions}
concludes the paper.

\section{Background}

\label{Section:Background}

The hybrid CRB provides a low bound on the MSE of any unbiased estimator for
an unknown parameter(s), where the parameters are partially deterministic and
partially random \cite{bell07}. Given a vector parameter $\mathbf{\theta
=}\left[  \mathbf{\theta}_{nr},\mathbf{\theta}_{r}\right]  ^{T}$, where
$\mathbf{\theta}_{nr}$ stands for the nonrandom parameter vector and
$\mathbf{\theta}_{r}$\ for a random parameter vector, its unbiased estimate
$\widehat{\mathbf{\theta}}$ satisfies the following inequality \cite{bell07}:%
\begin{equation}
E_{\mathbf{\theta}_{nr},\mathbf{\theta}_{r}}\left\{  \left(  \widehat{\theta
_{i}}-\theta_{i}\right)  \left(  \widehat{\theta_{i}}-\theta_{i}\right)
^{T}\right\}  \geq\left[  \mathbf{J}_{H}^{-1}\left(  \mathbf{\theta}%
_{nr},\mathbf{\theta}_{r}\right)  \right]  _{_{\text{\ }}i,i\text{\ }}\text{,}
\label{e:HCRB1}%
\end{equation}
where $\mathbf{J}_{H}\left(  \mathbf{\theta}\right)  $ is the hybrid Fisher
Information matrix (HFIM) expressed as%
\begin{equation}
\mathbf{J}_{H}\left(  \mathbf{\theta}_{nr},\mathbf{\theta}_{r}\right)
=\mathbf{J}_{D}+\mathbf{J}_{P}. \label{e:Jh_def}%
\end{equation}
The elements of the matrices $\mathbf{J}_{D}$\ and $\mathbf{J}_{P}$\ given by%
\begin{equation}
\left[  \mathbf{J}_{D}\right]  _{i,j}=-E_{\mathbf{\theta}_{r}\left\vert
\mathbf{\theta}_{nr}\right.  }\left\{  E_{\mathbf{r}\left\vert \mathbf{\theta
}_{nr},\mathbf{\theta}_{r}\right.  }\left\{  \frac{\partial\ln p\left(
\mathbf{r|\theta}_{nr},\mathbf{\theta}_{r}\right)  }{\partial\theta
_{i}\partial\theta_{j}}\right\}  \right\}  , \label{e:HCRBelements}%
\end{equation}
and
\begin{equation}
\left[  \mathbf{J}_{P}\right]  _{i,j}=-E_{\mathbf{\theta}_{r}\left\vert
\mathbf{\theta}_{nr}\right.  }\left\{  \frac{\partial^{2}\ln p\left(
\mathbf{\theta}_{r}\mathbf{|\theta}_{nr}\right)  }{\partial\theta_{i}%
\partial\theta_{j}}\right\}  ,\nonumber
\end{equation}
where $p\left(  \mathbf{r|\theta}_{nr},\mathbf{\theta}_{r}\right)  $ is the
conditional, joint probability density function (pdf) of the observations and
$p\left(  \mathbf{\theta}_{r}\mathbf{|\theta}_{nr}\right)  $ the conditional
joint pdf of $\mathbf{\theta}_{r}$. The matrix $\mathbf{J}_{D}$ represents the
contribution of the data and the matrix $\mathbf{J}_{P}$\ represents the
contribution of prior information.

The HCRB matrix is defined as%
\begin{equation}
\mathbf{HCRB}=\left[  \mathbf{J}_{H}\left(  \mathbf{\theta}_{nr}%
,\mathbf{\theta}_{r}\right)  \right]  ^{-1}. \label{e:HCRB2}%
\end{equation}

In cases in which the observation statistic is expressed in terms of $p\left(
\mathbf{r|\kappa}_{nr},\mathbf{\kappa}_{r}\right)  $, and the relationship
between the unknown parameters $\mathbf{\theta}_{nr},\mathbf{\theta}_{r}$ and
$\mathbf{\kappa}_{nr},\mathbf{\kappa}_{r}$ is given by $\kappa_{j}%
=f_{j}(\mathbf{\theta})$, the \textit{chain rule}, can be used to express
$\mathbf{J}_{H}\left(  \mathbf{\theta}_{nr},\mathbf{\theta}_{r}\right)  $ in
an alternative form \cite{KaySSPE93}:%
\begin{equation}
\mathbf{J}_{H}\left(  \mathbf{\theta}_{nr},\mathbf{\theta}_{r}\right)
=\mathbf{P}\left(  \mathbf{J}_{H}\left(  \mathbf{\kappa}_{nr},\mathbf{\kappa
}_{r}\right)  \right)  \mathbf{P}^{T}, \label{e:chain_rule}%
\end{equation}
where the elements of the matrix $\mathbf{P}$\ are given by $\left[
\mathbf{P}\right]  _{i,j}=\frac{\partial\kappa_{j}}{\partial\mathbf{\theta
}_{i}}$.

\section{HCRB\ with Phase Mismatch}

\label{Section:HCRB}

In this section, the HCRB is developed for target localization. A point target
is assumed with complex
reflectivity
$\vartheta=\vartheta_{\operatorname{Re}}+j\vartheta_{\operatorname{Im}}$,
located in a two dimensional plane at coordinates $X_{o}=\left(  x_{o}%
,y_{o}\right)  $. Consider a set of $M$ transmitting stations and $N$
receiving stations, widely distributed over a given geographical area, and
time and phase synchronized. A set of orthogonal waveforms is transmitted,
with the lowpass equivalents $s_{k}\left(  t\right)  ,$ $k=1,\ldots,M$, and
effective bandwidths $\beta$ \cite{Skolnik02}. The signals are narrowband in
the sense that for a carrier frequency of $f_{c}$, the narrowband signal
assumption implies $\beta^{2}/$\textbf{\ }$f_{c}^{2}\ll1$

In \cite{Godrich09}, perfect phase synchronization was assumed. In practice,
synchronization errors exists, modeled here as zero mean Gaussian random
variables with standard deviation $\sigma_{\Delta}^{2}$\ and denoted by
$\mathbf{\Delta\phi}=\left[  \Delta\phi_{t_{1}},\Delta\phi_{t_{2}}%
,...,\Delta\phi_{t_{M}},\Delta\phi_{r_{1}},\Delta\phi_{r_{2}},...,\Delta
\phi_{r_{N}}\right]  ^{T}$, where $\Delta\phi_{t_{k}}$ and $\Delta
\phi_{r_{\ell}}$ are phase errors at transmitting radar $k$ and receiving
radar $\ell$, respectively. The phase errors introduced by the different
stations are assumed to be statistically independent. The vector of unknown
parameters is defined by%
\begin{equation}
\mathbf{\theta}=\left[  \mathbf{\theta}_{nr},\mathbf{\theta}_{r}\right]  ^{T},
\label{e:tetha_def}%
\end{equation}
where $\mathbf{\theta}_{nr}=\left[  x_{o},y_{o},\vartheta_{\operatorname{Re}%
},\vartheta_{\operatorname{Im}}\right]  $ denotes the deterministic
unknowns\ and $\mathbf{\theta}_{r}=\mathbf{\Delta\phi}^{T}$ denotes the random unknowns.\ 

The estimation process is based on the signals observed at the receiving
sensors. The signal received at sensor $\ell$ is a superposition of the
transmitted signals,
reflected
from the target, and given by:%
\begin{equation}
r_{\ell}\left(  t\right)  =\overset{M}{\underset{k=1}{\sum}}\vartheta
s_{k}\left(  t-\tau_{\ell k}\right)  \eta_{\ell k}+n_{\ell}(t), \label{e:rl1}%
\end{equation}
where $\eta_{\ell k}$ accounts for the phase information and has the value of
$\eta_{\ell k}=\exp\left(  {-j2\pi f_{c}\tau_{\ell k}}\right)  \exp\left(
{-j}\left(  \Delta\phi_{t_{k}}+\Delta\phi_{r_{\ell}}\right)  \right)  $. The
noise $n_{\ell}\left(  t\right)  $ is assumed to be circularly symmetric,
zero-mean, complex Gaussian, spatially and temporally white with
autocorrelation function $\sigma_{n}^{2}\delta\left(  \tau\right)  $. The
propagation time, $\tau_{\ell k}$, is a sum of the time delays from station
$k$ to the target and from the target to station $\ell$, and may be expressed
as%
\begin{align}
\tau_{\ell k}  &  =\frac{1}{c}\left(  \sqrt{\left(  x_{tk}-x_{o}\right)
^{2}+\left(  y_{tk}-y_{o}\right)  ^{2}}\right. \label{e:tau_def}\\
&  \left.  +\sqrt{\left(  x_{r\ell}-x_{o}\right)  ^{2}+\left(  y_{r\ell}%
-y_{o}\right)  ^{2}}\right)  ,\nonumber
\end{align}
where $c$ denotes the speed of light, $\left(  x_{tk},y_{tk}\right)  $ denotes
the location of transmitting radar $k$ and $\left(  x_{r\ell},y_{r\ell
}\right)  $ denotes the location of receiving radar $\ell$. The following
vector notation is introduced: $\mathbf{\tau=}\left[  \tau_{11},\tau
_{12},...,\tau_{\ell k},...,\tau_{NM}\right]  ^{T}$.

The received signals are separated at the receiver by exploiting the
orthogonality between the transmitted waveforms. The signal in (\ref{e:rl1})
is defined as a function of the time of arrival, $\tau_{\ell k}$, the
reflectivity
value $\vartheta$, and the phase mismatch $\mathbf{\Delta\phi}$. The vector of
unknown parameters for the observations $r_{\ell}\left(  t\right)  $ is
expressed as a function of the time delays $\mathbf{\tau}$\ rather than a
function of the unknown location $\left(  x_{o},y_{o}\right)  $\ (as seen in
(\ref{e:tau_def})); i.e., the vector of unknown parameters is denoted by
$\mathbf{\kappa=}\left[  \mathbf{\kappa}_{nr},\mathbf{\kappa}\right]  ^{T}$,
with $\mathbf{\kappa}_{nr}\mathbf{=}\left[  \mathbf{\tau}^{T},\vartheta
_{\operatorname{Re}},\vartheta_{\operatorname{Im}}\right]  $\ and
$\mathbf{\kappa}_{r}=\mathbf{\Delta\phi}^{T}$. The following notation is
defined for later use: $\mathbf{r}=\left[  r_{1}\left(  t\right)
,\ldots,r_{N}\left(  t\right)  \right]  ^{T}$, $Q=MN$, $L=M+N.$

In order to derive the HFIM\ given in (\ref{e:Jh_def}) and
(\ref{e:HCRBelements}), the conditional joint pdf $p\left(  \mathbf{r}%
|\mathbf{\kappa}\right)  $ is required. For the signal model given in
(\ref{e:rl1}), the conditional joint pdf of the observations (time samples at
multiple receive antennas) parametrized by the unknown parameters vector
$\mathbf{\kappa}$, is then%
\begin{align}
p\left(  \mathbf{r}|\mathbf{\kappa}\right)   &  \varpropto\exp
\label{e:rl1_pdf}\\
&  \left\{  -\frac{1}{\sigma_{n}^{2}}\overset{N}{\underset{\ell=1}{\sum}}%
\int_{T}\left\vert r_{\ell}(t)-\overset{M}{\underset{k=1}{\sum}}\vartheta
s_{k}\left(  t-\tau_{\ell k}\right)  \eta_{\ell k}\right\vert ^{2}dt\right\}
.\nonumber
\end{align}
\newline The observation is given as a function of $\mathbf{\kappa}$.
Therefore, the matrix $\mathbf{P}$, defined following (\ref{e:chain_rule}),
needs to be derived. The relation given in (\ref{e:tau_def}) is used,
resulting in%
\begin{equation}
\mathbf{P}=\left[
\begin{array}
[c]{cc}%
\mathbf{D}_{2\times Q}^{T} & \mathbf{0}\\
\mathbf{0} & \mathbf{I}_{\left(  2+L\right)  \times\left(  2+L\right)  }%
\end{array}
\right]  , \label{e:P}%
\end{equation}
with%
\begin{equation}
\mathbf{D}={\small -}\frac{1}{c}\left[
\begin{array}
[c]{cc}%
\cos\alpha_{1}+\cos\gamma_{2} & \sin\alpha_{2}+\sin\gamma_{2}\\
\vdots & \vdots\\
\cos\alpha_{M}+\cos\gamma_{N} & \sin\alpha_{M}+\sin\gamma_{N}%
\end{array}
\right]  ^{T}, \label{e:D}%
\end{equation}
where $\alpha_{k}$ is the bearing angle of the transmitting sensor $k$ to the
target, measured with respect to the $x$ axis, and $\gamma_{\ell}$ is the
bearing angle of the receiving radar $\ell$ to the target, measured with
respect to the $x$ axis.

Using the conditional pdf $p\left(  \mathbf{r}|\mathbf{\kappa}\right)  $ in
(\ref{e:rl1_pdf}) and the Gaussian distribution of the phase errors, the
HFIM\ $\mathbf{J}_{H}\left(  \mathbf{\kappa}\right)  $, defined by
(\ref{e:Jh_def}) and (\ref{e:HCRBelements}), is derived in Appendix
\ref{Section:appendixA}, resulting in%
\begin{equation}
\mathbf{J}_{H}\left(  \mathbf{\kappa}_{nr},\mathbf{\kappa}_{r}\right)
=\left[
\begin{array}
[c]{cc}%
\mathbf{R}_{\tau} & \mathbf{G}\\
\mathbf{G}^{T} & \mathbf{H}%
\end{array}
\right]  , \label{e:Jh-mimo}%
\end{equation}
where matrices $\mathbf{G}$ and $\mathbf{H}$ are defined by%
\begin{equation}
\mathbf{G}=\left[
\begin{array}
[c]{cc}%
\mathbf{F}_{\tau\vartheta} & \mathbf{F}_{\tau\Delta}%
\end{array}
\right]  _{Q\times\left(  2+L\right)  }, \label{e:G}%
\end{equation}
and
\begin{equation}
\mathbf{H}=\left[
\begin{array}
[c]{cc}%
\mathbf{\Sigma}_{\vartheta} & \mathbf{F}_{\vartheta\Delta}\\
\mathbf{F}_{\vartheta\Delta}^{T} & \mathbf{\Sigma}_{\Delta}+\frac{1}%
{\sigma_{\Delta}^{2}}\mathbf{I}%
\end{array}
\right]  _{\left(  2+L\right)  \times\left(  2+L\right)  },
\end{equation}
and the other submatrices in (\ref{e:Jh-mimo}), (\ref{e:G})\ and (\ref{e:H})
are defined and derived in Appendix \ref{Section:appendixA} (see
(\ref{e:FtauTau}), (\ref{e:Freflect}), (\ref{e:Fdelta}) and
(\ref{e:CrossMatrix})). Applying (\ref{e:P}) and (\ref{e:Jh-mimo}) in
(\ref{e:chain_rule}) yields%

\begin{equation}
\mathbf{J}_{H}\left(  \mathbf{\theta}\right)  =\left[
\begin{array}
[c]{cc}%
\mathbf{DR}_{\tau}\mathbf{D}^{T} & \mathbf{DG}\\
\mathbf{G}^{T}\mathbf{D}^{T} & \mathbf{H}%
\end{array}
\right]  . \label{e:Jh-mimo2}%
\end{equation}
The HCRB for the unknown parameters $\left(  x_{o},y_{o}\right)  $ may be
derived from (\ref{e:Jh-mimo2}), applying the relation given in (\ref{e:HCRB2}%
) \cite{Godrich09}:%
\begin{equation}
\mathbf{HCRB}\left(  x_{o},y_{o}\right)  =\left[  \mathbf{DR}_{\tau}%
\mathbf{D}^{T}-\mathbf{DGH}^{-1}\mathbf{G}^{T}\mathbf{D}^{T}\right]
_{2\times2}^{-1}. \label{e:HCRBxy}%
\end{equation}
To find the closed-form solution to $\mathbf{HCRB}\left(  x_{o},y_{o}\right)
$, the matrix $\mathbf{H}^{-1}$ is expressed using the formula for the inverse
of a partitioned matrix \cite{Horn90}:%
\begin{align}
\left[  \mathbf{H}^{-1}\right]  _{11}  &  =\left[  \mathbf{\Sigma}_{\vartheta
}-\mathbf{F}_{\vartheta\Delta}\mathbf{A}_{\Delta}^{-1}\mathbf{F}%
_{\vartheta\Delta}^{T}\right]  ^{-1}\label{e:H11}\\
\left[  \mathbf{H}^{-1}\right]  _{22}  &  =-\left[  \mathbf{F}_{\vartheta
\Delta}^{T}\mathbf{\Sigma}_{\vartheta}^{-1}\mathbf{F}_{\vartheta\Delta}%
^{T}-\mathbf{A}_{\Delta}\right]  ^{-1} \label{e:H22}%
\end{align}
and
\begin{equation}
\left[  \mathbf{H}^{-1}\right]  _{12}=\left[  \mathbf{H}^{-1}\right]
_{21}^{T}=\mathbf{\Sigma}_{\vartheta}^{-1}\mathbf{F}_{\vartheta\Delta}\left[
\mathbf{F}_{\vartheta\Delta}^{T}\mathbf{\Sigma}_{\vartheta}^{-1}%
\mathbf{F}_{\vartheta\Delta}-\mathbf{A}_{\Delta}\right]  ^{-1} \label{e:H21}%
\end{equation}
where $\mathbf{A}_{\Delta}=\left(  \mathbf{\Sigma}_{\Delta}+\frac{1}%
{\sigma_{\Delta}^{2}}\mathbf{I}\right)  $. The term $\left[  \mathbf{\Sigma
}_{\vartheta}-\mathbf{F}_{\vartheta\Delta}^{-1}\mathbf{A}_{\Delta}%
^{-1}\mathbf{F}_{\vartheta\Delta}^{T}\right]  ^{-1}$ in (\ref{e:H11}), is
transformed based on the formula for the inverse of a matrix $\mathbf{B}$ of
the form $\mathbf{B=A+XRY}$, given in \cite{Horn90}. Following some additional
matrix manipulations, the HCRB for the location MSE can be expressed as%
\begin{align}
\mathbf{HCRB}\left(  x_{o},y_{o}\right)   &  =\mathbf{J}_{F}^{-1}%
\mathbf{+}\left[  \mathbf{J}_{F}-\mathbf{J}_{F}\mathbf{P}_{\Delta}%
^{-1}\mathbf{J}_{F}\right]  ^{-1}\label{e:HCRBex}\\
&  =\mathbf{CRB}_{o}\left(  x_{o},y_{o}\right)  +\Delta\mathbf{CRB,}\nonumber
\end{align}
where $\mathbf{CRB}_{o}\left(  x_{o},y_{o}\right)  =\mathbf{J}_{F}^{-1}$ is
the CRB\ with no phase mismatch, and $\Delta\mathbf{CRB=}\left[
\mathbf{J}_{F}-\mathbf{J}_{F}\mathbf{P}_{\Delta}^{-1}\mathbf{J}_{F}\right]
^{-1}$ represents the increment in the bound due to phase synchronization
errors. The matrices $\mathbf{J}_{F}$ and $\mathbf{P}_{\Delta}$\ are defined by%

\[
\mathbf{J}_{F}=\mathbf{DR}_{\tau}\mathbf{D}^{T}-\mathbf{D\mathbf{F}%
_{\tau\vartheta}\mathbf{\Sigma}_{\vartheta}^{-1}\mathbf{F}_{\tau\vartheta}%
^{T}D}^{T},
\]
and
\begin{align}
\mathbf{P}_{\Delta}  &  =\mathbf{DF}_{\tau\vartheta}\mathbf{\Sigma}%
_{\vartheta}^{-1}\mathbf{F}_{\vartheta\Delta}\mathbf{R}_{\Delta}%
^{-1}\mathbf{F}_{\vartheta\Delta}^{T}\mathbf{\Sigma}_{\vartheta}%
^{-1}\mathbf{\mathbf{F}_{\tau\vartheta}^{T}D}^{T}\label{e:Pdelta1}\\
&  -2\operatorname{Re}\left\{  \mathbf{D\mathbf{F}_{\tau\Delta}R}_{\Delta
}^{-1}\mathbf{F}_{\vartheta\Delta}^{T}\mathbf{\Sigma}_{\vartheta}%
^{-1}\mathbf{\mathbf{F}_{\tau\vartheta}^{T}D}^{T}\right\} \nonumber\\
&  +\mathbf{D\mathbf{F}_{\tau\Delta}R}_{\Delta}^{-1}\mathbf{\mathbf{F}%
_{\tau\Delta}^{T}D}^{T},\nonumber
\end{align}
and the matrix $\mathbf{R}_{\Delta}^{-1}$\ can be calculated recursively using
the formula for the inverse of the sum of matrices \cite{Miller81}, resulting
in
\begin{equation}
\mathbf{R}_{\Delta}^{-1}=\left[
\begin{array}
[c]{cc}%
\lambda_{1}\mathbf{I}_{M\times M}+\frac{N\lambda_{1}^{2}}{M\left(
1-N\lambda_{1}\right)  }\mathbf{11}^{T} & \mathbf{0}\\
\mathbf{0} & \lambda_{2}\mathbf{I}_{N\times N}+\frac{M\lambda_{2}^{2}%
}{N\left(  1-M\lambda_{2}\right)  }\mathbf{11}^{T}%
\end{array}
\right]  , \label{e:RdeltaInv}%
\end{equation}
where $\mathbf{1}=\left[  1,1,,...,1\right]  ^{T}$ and the terms $\lambda_{1}$
and $\lambda_{2}$ are
\begin{equation}%
\begin{array}
[c]{cc}%
\lambda_{1}=\left(  N+\frac{1}{2\operatorname*{snr}\sigma_{\Delta}^{2}%
}\right)  ^{-1}\text{ and} & \lambda_{2}=\left(  M+\frac{1}%
{2\operatorname*{snr}\sigma_{\Delta}^{2}}\right)  ^{-1}%
\end{array}
. \label{e:lambda}%
\end{equation}
Calculating the explicit value of $\Delta\mathbf{CRB}$, we get%
\begin{equation}
\Delta\mathbf{CRB}=\left[  \mathbf{J}_{F}-\mathbf{J}_{F}\left(  \mu
_{0}\overset{3}{\underset{m=1}{\sum}}k_{m}\mathbf{B}_{m}\right)
^{-1}\mathbf{J}_{F}\right]  _{2\times2}^{-1}, \label{e:PdeltaExp}%
\end{equation}
where the constants\ $\mu_{m}$, $m=0,...,3$ are functions of the phase
synchronization error variance $\sigma_{\Delta}^{2}$ (through $\lambda_{1} $
and $\lambda_{2}$, defined in (\ref{e:lambda})) and the number of transmitting
and receiving radars $M$ and $N$, as follows:%
\begin{align}
\mu_{0}  &  =\frac{8\pi^{2}\left(  f_{c}^{2}+\beta^{2}\right)
\operatorname*{snr}}{c^{2}},\label{e:ks}\\
\mu_{1}  &  =\lambda_{1}/M+\lambda_{2}/N,\nonumber\\
\mu_{2}  &  =\lambda_{1}\left(  N^{2}/M\right)  \text{ and\ }k_{3}=\lambda
_{2}\left(  M^{2}/N\right)  .\nonumber
\end{align}
The matrices $\mathbf{B}_{m}$, $m=1,2,3$\ depend on the geographical layout of
the radars with respect to the target location:%
\[
\mathbf{B}_{m}=\left[
\begin{array}
[c]{cc}%
\left[  \mathbf{D}_{m}\mathbf{1}\right]  _{1,1}^{2} & \left[  \mathbf{D}%
_{m}\mathbf{1}\right]  _{1,1}\left[  \mathbf{D}_{m}\mathbf{1}\right]  _{2,1}\\
\left[  \mathbf{D}_{m}\mathbf{1}\right]  _{1,1}\left[  \mathbf{D}%
_{m}\mathbf{1}\right]  _{2,1} & \left[  \mathbf{D}_{m}\mathbf{1}\right]
_{2,1}^{2}%
\end{array}
\right]  ,
\]
using the following $\mathbf{D}_{m}$ matrices:
\begin{align}
\mathbf{D}_{1}  &  ={\small c}\mathbf{D}\label{e:Dmulti}\\
\mathbf{D}_{2}  &  =\left[
\begin{array}
[c]{cc}%
\cos\alpha_{1} & \sin\alpha_{2}\\
\vdots & \vdots\\
\cos\alpha_{M} & \sin\alpha_{M}%
\end{array}
\right]  _{M\times2}^{T},\nonumber
\end{align}
and%
\[
\mathbf{D}_{3}=\left[
\begin{array}
[c]{cc}%
\cos\gamma_{2} & \sin\gamma_{2}\\
\vdots & \vdots\\
\cos\gamma_{N} & \sin\gamma_{N}%
\end{array}
\right]  _{N\times2}^{T}.
\]
The expression for the HCRB as given in (\ref{e:HCRBex}), offers an
interesting observation on the effects of phase errors on the target
localization MSE. First, it is apparent that the HCRB may be expressed as the
sum of the CRB with no phase error and a term dependent on the statistics of
the phase errors. This term is a function of the sensors location with respect
to the target, through the matrices $\mathbf{B}_{m}$, and the system
parameters (SNR, phase errors variance $\sigma_{\Delta}^{2}$ and the number of
mismatched transmitting and receiving radars) through the coefficients
$\mu_{m}$. The manner in which the number of radars, their spread and the
phase synchronization error variance affect the performance is not readily
understood from (\ref{e:PdeltaExp}). For this reason, numerical examples are
employed in the next section to gain some insight into the relationships
between system parameters and performance degradation.

\section{Numerical Analysis}

\label{Section:NumericalAnalysis}

We have evaluated the HCRB expression given in (\ref{e:HCRBex})\ numerically
using the following example: $M=11,$ $N=9$ and $\sigma_{\Delta}^{2}=\left[
0,0.0001,0.0005,0.001,0.005,0.01,0.05\right]  $, where $\sigma_{\Delta}^{2}$
is expressed in $\left(  rad^{2}\right)  $. The $\mathbf{HCRB}\left(
x_{o},y_{o}\right)  $ is drawn in Figure 1. As $\sigma_{\Delta}^{2}%
$\ increases beyond a specific value, the additional CRB term $\Delta
\mathbf{CRB}$ dominates the performance and the curve. For high phase error
levels, the performance degradation starts at lower SNRs. For small phase
errors, localization accuracy is not undermined by the phase mismatch, and the
$\mathbf{HCRB}\left(  x_{o},y_{o}\right)  $\ curve follows the $\mathbf{CRB}%
_{o}\left(  x_{o},y_{o}\right)  $ closely.%

\begin{figure}[ptb]
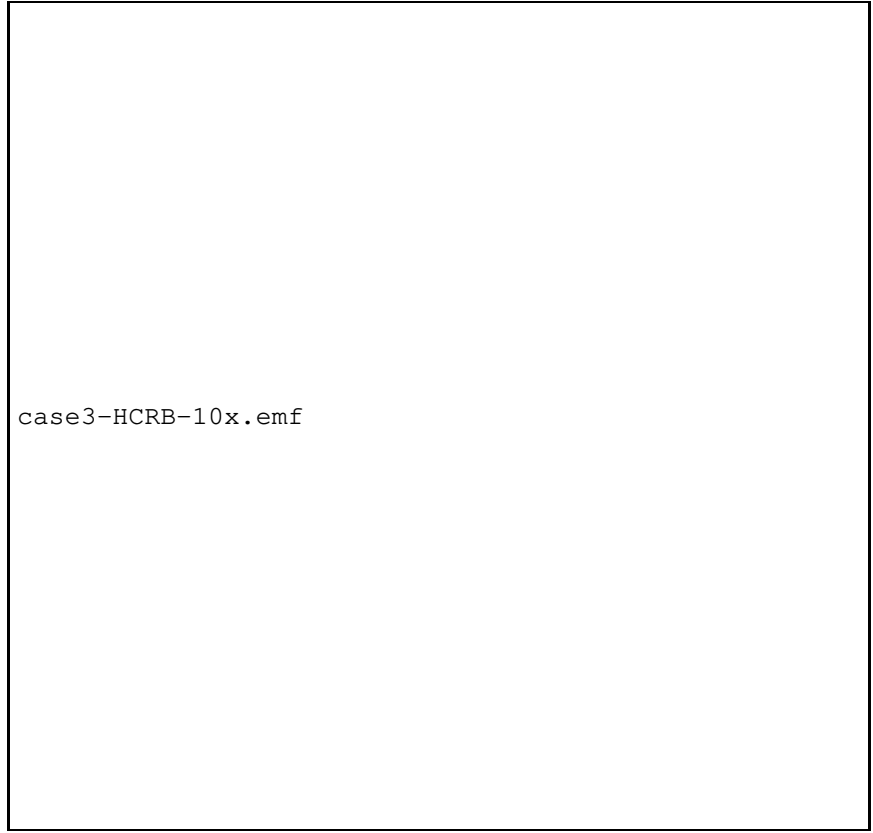
%
\centering
\ifcase\msipdfoutput
\includegraphics[
natheight=10.027500in,
natwidth=10.430500in,
height=4.3396in,
width=4.5126in
]%
{case3-HCRB-10x.emf}%
\else
\includegraphics[
natheight=4.339600in,
natwidth=4.512600in,
height=4.3396in,
width=4.5126in
]%
{C:/Users/Hana/Desktop/My Papers/CAMSAP_2009/graphics/case3-HCRB-10x__1.pdf}%
\fi
\caption{HCRB for M=11 and N=9. The blue line represent the CRB value with no
phase errors. $\sigma_{\Delta}^{2}$ range from 0 to 0.05.}%
\label{f:1}%
\end{figure}

For a given system, the tolerated $\left[  \sigma_{\Delta}^{2}\right]  _{\max
}$ may be determined by solving $\Delta\mathbf{CRB\left(  \left[
\sigma_{\Delta}^{2}\right]  _{\max}\right)  \preceq CRB}_{o}\left(
x_{o},y_{o}\right)  $. This value can serve as a design goal in the system
phase calibration. For a given phase synchronization error variance
$\sigma_{\Delta}^{2}$, the expression $\Delta\mathbf{CRB\left(  \sigma
_{\Delta}^{2}\right)  }$ gives the localization accuracy penalty.

\section{Conclusions}

\label{Section:Conclusions}

MIMO radar with coherent processing exploits the signal phase measured at the
receive antennas to generate high resolution target location estimation. To
take advantage of this scheme, full phase synchronization is required among
all participating radars. In practice, inevitable phase synchronization
errors
reflect
on the system localization performance. In this paper, a closed-form
expression of the HCRB of target localization has been derived, capturing the
impact of the phase synchronization errors on the achievable target
localization accuracy. In particularly it has been shown that the HCRB can be
expressed as a sum of the CRB with no phase error and a term that represents
the phase error penalty. The latter has been shown to be a function of the
sensors geometry, SNR, and the number of transmitting and receiving radars in
addition to the phase error MSE. As phase synchronization over distributed
platform is a complex operation and phase errors are unavoidable, the HCRB
offers valuable information at the system design level. For a given phase
error MSE, the HCRB may be used to derive the attainable target localization
accuracy. Otherwise, for a given system performance goal on localization
accuracy, the HCRB provides with an upper bound on the necessary phase error
MSE values.

\appendices
%

\section{Derivation of the J(D) matrix}%

\label{Section:appendixA}

In this appendix, we develop the elements of the matrix $\mathbf{J}_{D}\left(
\mathbf{\kappa}\right)  $, i.e. $\left[  \mathbf{J}_{D}\left(  \mathbf{\kappa
}\right)  \right]  _{i,,j}=-E_{\mathbf{\kappa}_{r}\left\vert \mathbf{\kappa
}_{nr}\right.  }\left\{  E_{\mathbf{r}\left\vert \mathbf{\kappa}\right.
}\left[  \frac{\partial^{2}\ln\left(  p\left(  \mathbf{r}\left\vert
\mathbf{\kappa}\right.  \right)  \right)  }{\partial\kappa_{i}\partial
\kappa_{j}}\right]  \right\}  $, based on the conditional pdf in
(\ref{e:rl1_pdf}).The diagonal submatrix $\mathbf{R}_{\tau}$ is derived as
follows:%
\begin{align}
\left[  \mathbf{R}_{\tau}\right]  _{i,j}  &  =-E_{\mathbf{\kappa}%
_{r}\left\vert \mathbf{\kappa}_{nr}\right.  }\left\{  E_{\mathbf{r}\left\vert
\mathbf{\kappa}\right.  }\left[  \frac{\partial^{2}\ln\left(  p\left(
\mathbf{r}|\mathbf{\kappa}\right)  \right)  }{\partial{\tau_{\ell k}}%
\partial{\tau_{\ell^{\prime}k^{\prime}}}}\right]  \right\} \label{e:FtauTau}\\
&  =\sigma_{n}^{2}\operatorname{Re}\left\{  \left\vert \vartheta\right\vert
^{2}\left[
\begin{array}
[c]{c}%
\frac{\partial^{2}}{\partial{\tau_{\ell k}}\partial{\tau_{\ell^{\prime
}k^{\prime}}}}\int\eta_{\ell k}\eta_{\ell^{\prime}k^{\prime}}^{\ast}\\
\cdot{\small s}_{k}\left(  t-{\tau_{\ell k}}\right)  {\small s}_{k^{\prime}%
}^{\ast}\left(  t-{\tau_{\ell^{\prime}k}^{\prime}}\right)
\end{array}
\right]  \right\}  ,\nonumber
\end{align}
and
\[
\mathbf{R}_{\tau}=8\pi^{2}\left(  f_{c}^{2}+\beta^{2}\right)
\operatorname*{snr}\mathbf{I}_{Q\times Q},\text{ }%
\]
where $\operatorname*{snr}=\left\vert \vartheta\right\vert ^{2}/\sigma_{n}%
^{2}$, and the following notation is used:%
\begin{equation}%
\begin{array}
[c]{cc}%
i=\left[  (\ell-1)\ast M+k\right]  \ \text{\ and} & j=\left[  (\ell^{\prime
}-1)\ast M+k^{\prime}\right]  ,\\
\text{\ }k,k^{\prime}=1,..,M, & \ell,\ell^{\prime}=1,..,N,
\end{array}
\text{.} \label{e:indexing}%
\end{equation}
The elements of the matrix $\mathbf{\Sigma}_{\vartheta}$ are given by
\begin{equation}
\left[  \mathbf{\Sigma}_{\vartheta}\right]  _{1,1}=\frac
{2MN\operatorname*{snr}}{\left\vert \vartheta\right\vert ^{2}}=\left[
\mathbf{\Sigma}_{\vartheta}\right]  _{2,2}, \label{e:Freflect}%
\end{equation}
and%
\begin{equation}
\left[  \mathbf{\Sigma}_{\vartheta}\right]  _{1,2}=0=\left[  \mathbf{\Sigma
}\right]  _{2,1},\nonumber
\end{equation}
and the elements of the matrix $\mathbf{\Sigma}_{\Delta}$ are given by
\begin{equation}
\mathbf{\Sigma}_{\Delta}=2\operatorname*{snr}\left[
\begin{array}
[c]{cc}%
N\mathbf{I}_{M\times M} & \left(  \mathbf{11}^{T}\right)  _{M\times N}\\
\left(  \mathbf{11}^{T}\right)  _{N\times M} & M\mathbf{I}_{N\times N}%
\end{array}
\right]  _{L\times L}. \label{e:Fdelta}%
\end{equation}
The off-diagonal submatrices are as follows:
\begin{equation}%
\begin{array}
[c]{c}%
\mathbf{F}_{\tau\vartheta_{Q\times2}}=\frac{4\pi f_{c}}{\sigma_{n}^{2}}\left[
\begin{array}
[c]{cc}%
\vartheta_{\operatorname{Im}}\mathbf{1}_{Q\times1} & -\vartheta
_{\operatorname{Re}}\mathbf{1}_{Q\times1}%
\end{array}
\right]  ,\\
\mathbf{F}_{\tau\Delta_{Q\times L}}=4\pi f_{c}\operatorname*{snr}\left[
\left.
\begin{array}
[c]{c}%
\mathbf{I}_{M\times M}\\
\vdots\\
\mathbf{I}_{M\times M}%
\end{array}
\right\vert
\begin{array}
[c]{c}%
\Pi\left(  1\right) \\
\vdots\\
\Pi\left(  N\right)
\end{array}
\right]  _{Q\times L},\\
\Pi\left(  \ell\right)  =\left[
\begin{array}
[c]{ccc}%
\mathbf{0}_{N\times\left(  \ell-1\right)  } & \mathbf{1}_{N\times1} &
\mathbf{0}_{N\times\left(  N-\ell-1\right)  }%
\end{array}
\right]  _{N\times N}%
\end{array}
\label{e:CrossMatrix}%
\end{equation}
and%
\begin{equation}%
\begin{array}
[c]{c}%
\mathbf{F}_{\vartheta\Delta_{2\times L}}=\frac{2\operatorname*{snr}%
}{\left\vert \vartheta\right\vert ^{2}}\left[
\begin{array}
[c]{cc}%
\vartheta_{\operatorname{Im}}N\mathbf{1}_{1\times M}^{T} & \vartheta
_{\operatorname{Im}}M\mathbf{1}_{1\times N}^{T}\\
-\vartheta_{\operatorname{Re}}N\mathbf{1}_{1\times M}^{T} & -\vartheta
_{\operatorname{Re}}M\mathbf{1}_{1\times N}^{T}%
\end{array}
\right]  .
\end{array}
\end{equation}


\begin{thebibliography}{99}                                                                                               %


\bibitem {Fishler04}E. Fishler, A. M. Haimovich, R. S. Blum, L. Cimini, D.
Chizhik, and R. Valenzuela, "MIMO radar: An idea whose time has
come,\textquotedblright\ in \emph{Proc. of the 2004 IEEE Int. Conf. on Radar,}
Philadelphia, April 2004, pp. 71--78

\bibitem {HaimovichBook}H. Godrich, A. M. Haimovich and R. S. Blum, "Concepts
and Applications of a MIMO Radar System with Widely Separated Antennas," book
chapter in \emph{MIMO Radar Signal Processing}, John Wiley 2008.

\bibitem {Haimovich08}A. Haimovich, R. Blum, and L. Cimini, "MIMO radar with
widely separated antennas," \emph{IEEE Signal Proc. Magazine}, Vol. 25,
January 2008, pp. 116 - 129.

\bibitem {Jian07}J. Li, and P. Stoica, "MIMO radar with colocated antennas,"
\emph{IEEE Signal Proc. Magazine}, Vol. 24, September 2007, pp. 106--114.

\bibitem {Godrich09}H. Godrich, A. M. Haimovich, and R. S. Blum, "Target
localization accuracy gain in MIMO radar based system,\textquotedblright%
\ submitted to \emph{IEEE Trans. on Information Theory.}

\bibitem {Poor}H. V. Poor, \emph{An Introduction to Signal Detection and
Estimation}, New York; Springer, 2nd ed, 1994

\bibitem {bell07}H. L. Van Trees, and K. L. Bell, \emph{Bayesian Bounds for
Parameter Estimation and Nonlinear Filtering/Tracking}, New York;
Wiley-Interscience, 2007.

\bibitem {Rockah87}Y. Rockah, H. Messer, and P. M. Schultheiss, "Localization
performance of arrays subject to phase errors," \emph{IEEE Trans. Acoust.,
Speech, Signal Proc.,} Vol. ASSP-35, March 1987, pp. 286-299.

\bibitem {Messer88}Y. Rockah, and P. M. Schultheiss, "Array shape calibration
using sources in unknown locations - part I: Far-field sources," \emph{IEEE
Trans. Acoust., Speech, Signal Proc.,} Vol. ASSP-35, March 1987, pp. 286-299.

\bibitem {Skolnik02}M. Skolnik,\emph{\ Introduction to Radar Systems}, New
York: McGraw-Hill, 3rd, 2002.

\bibitem {KaySSPE93}S. M. Kay, \emph{Fundementals of Statistical Signal
Processing: Estimation Theory,} vol. 1, Upper Saddle River, NJ: Parentice Hall
PTR, 1st ed., 1993.

\bibitem {Horn90}R. A. Horn, and C. R. Johnson, \emph{Matrix Analysis,}
Cambridge, UK:Cambridge University Press, 1990.

\bibitem {Miller81}K. S. Miller, \textquotedblleft On the inverse of the sum
of matrices,\textquotedblright\ Mathematics Magazine, Vol. 54, No. 2, 1981,
pp. 67-72.
\end{thebibliography}
\end{document}